\documentclass[11pt,twoside]{article}

\usepackage{asp2006}
\usepackage{epsf}

\begin{document}
\title{The Properties and Gaseous Environments of Powerful 
Classical Double Radio Galaxies}   
\author{Ruth A. Daly\altaffilmark{1}, C. P. O'Dea\altaffilmark{2},  
P. Kharb\altaffilmark{2}, S. A. Baum\altaffilmark{3}, 
Kenneth A. Freeman\altaffilmark{1}, and Matthew P. Mory\altaffilmark{1} }   
\altaffiltext{1}{Department of Physics, Penn State University, Berks Campus,
Reading, PA 19610}
\affil{}    
\altaffiltext{2}{Department of Physics, 
Rochester Institute of Technology, Rochester, NY 14623 }
\altaffiltext{3}{Center for Imaging Science, Rochester Institute of Technology,
Rochester, NY 14623}

\begin{abstract} 
The properties of a sample of 31 very powerful classical double
radio galaxies with redshifts between zero and 1.8 are studied.
The source velocities, beam powers, ambient gas densities,
total lifetimes, and total outflow energies are presented
and discussed. The rate of growth of each side of each
source were obtained using a spectral aging analysis.
The beam power and ambient gas density 
were obtained by applying the
strong shock jump conditions to the ends of each side of the
source.  The total outflow lifetime was obtained by applying
the power-law relationship between the beam power and the total
source lifetime derived elsewhere for sources of this type, and
the total outflow energy was obtained by combining the beam
power and the total source lifetime. 

Composite profiles were constructed by combining results obtained from 
each side of each source.  The composite profiles indicate that the 
ambient gas density falls with distance from the central engine.
The source velocities, beam powers, total lifetimes, and total energies
seem to be independent of radio source size.  This is consistent with 
the standard model in which each source grows at a roughly constant
rate during which time the central engine puts out a roughly constant 
beam power.  The fact that the total source lifetimes and energies
are independent of radio source size indicates that the 
sources are being sampled at random times during their lifetimes.

\end{abstract}



\section{Introduction}
\begin{figure}
\plotone{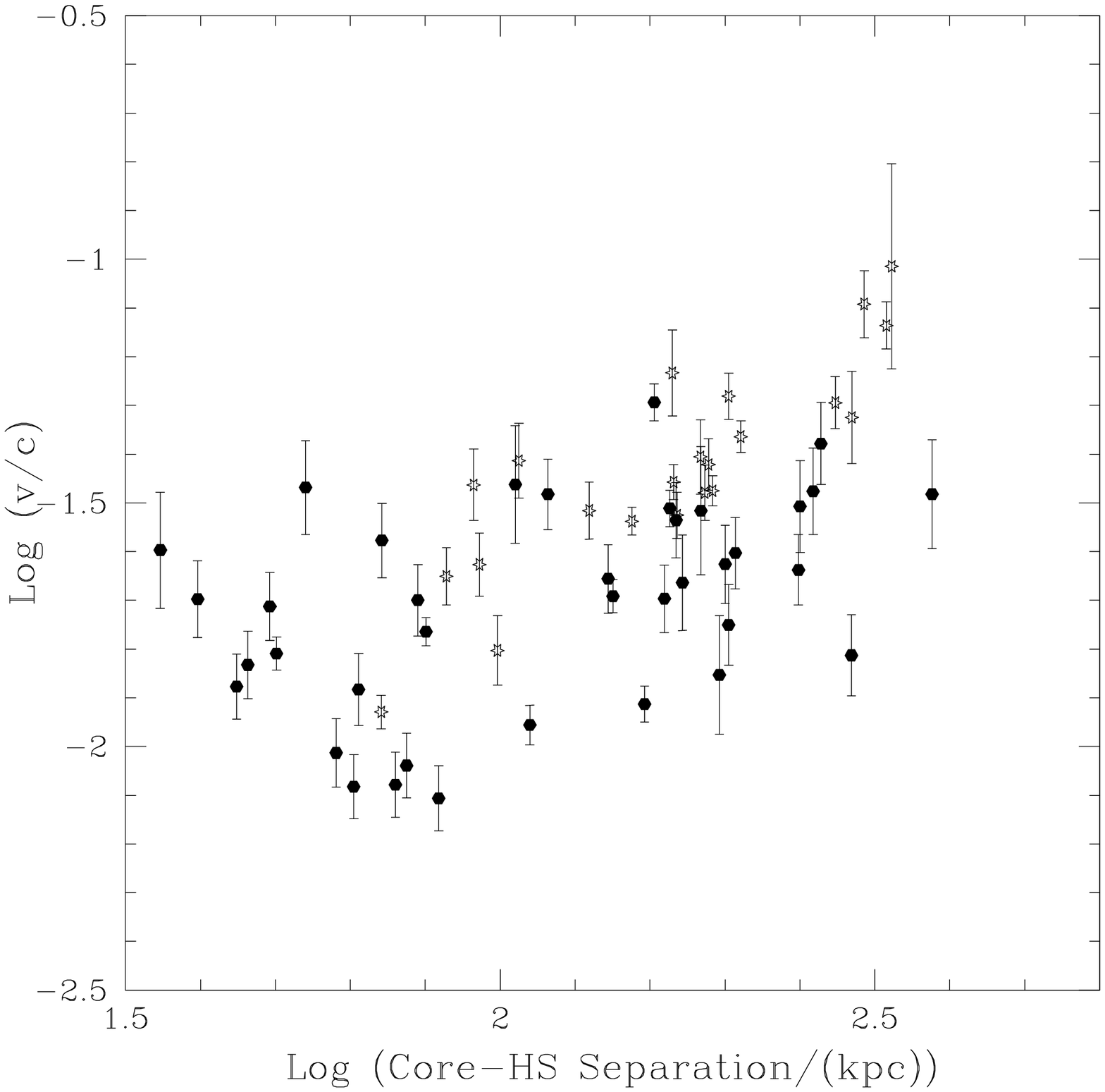}
\caption{The rate of growth or lobe propagation velocity of each side of each 
source as a function of core-hotspot separation for a 
magnetic field strength 
that is one quarter of the minimum energy value. The 
11 radio galaxies from \citet{K08} are indicated by open stars and 
20 sources from \citet{LMS89} and \citet{LPR92}    
are indicated by filled circles. Here, and 
in all of the figures, each
side of each source is represented by one data point. }
\end{figure}

Powerful extended radio 
galaxies are beacons that can be observed to very high 
redshift. A special and select subset of radio sources are 
studied here: very powerful classical double radio sources
with cigar-like bridge shapes.  \citet{LMS89} 
determined that sources of this type 
are among the most powerful FRII radio sources, and that these sources 
have quite regular radio bridge structure indicating that each 
side of each source is growing at a rate that is well into the supersonic 
regime and has negligible backflow. 
It follows that the equations governing strong shocks may be 
applied to these systems. Strong shock physics is clean and simple, and 
makes these sources ideal candidates for detailed study and 
analysis \citep[see][]{DY02}. 
Thus, only powerful extended radio sources with  
178 MHz radio powers greater than $3 \times 
10^{26}~h^{-2} \hbox{W Hz}^{-1} \hbox{sr}^{-1}$, where $H_0 = 100 ~h \hbox{ 
km s}^{-1} \hbox{Mpc}^{-1}$ are included in this study, and 
only radio galaxies are included so as to 
minimize projection effects, which are likely to be much more important 
for radio loud quasars. The sample includes 11 radio galaxies with
bright bridge emission presented
by \citet{K08}, six sources presented by \citet{GDW00}, 
and fourteen sources observed by \citet{LMS89} 
and \citet{LPR92}, which were 
also studied by \citet{WDW97}, for a total of 31 sources.
A standard cosmological model with 
$H_0 = 70 \hbox{ km s}^{-1} \hbox{Mpc}^{-1}$, $\Omega_m = 0.3$, 
and $\Omega_{\Lambda}=0.7$ is used throughout.
\begin{figure}
\plotone{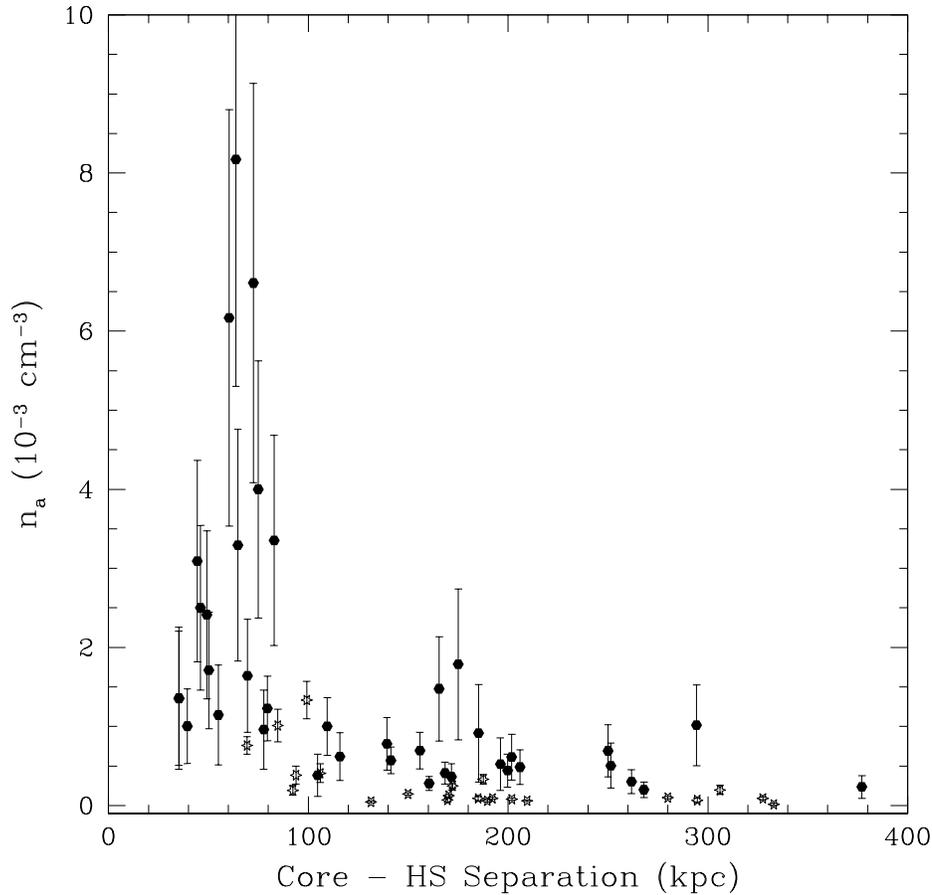}
\caption{As in Figure 1, but for the ambient gas density at 
the extremities of each side 
of each source.}
\end{figure}

\section{Results}
The work presented here is 
described in greater detail by \citet{O07}.  
A spectral aging analysis was carried out on eleven radio galaxies
presented by \citet{K08}, 
and combined with information obtained previously for twenty radio galaxies.
The velocities are shown in Figure 1 as a function of core-hotspot
separation. Two points are obtained for each source; one from each
side of the source.  The strong 
shock equation relating the ambient gas density $\rho$ to the bridge 
pressure $P$ and the rate of growth of the source $v$ ($\rho_a \propto 
P/v^2$), and that relating the beam power $L_j$ to these parameters and 
the bridge width $a$ ($L_j \propto P v a^2$) were used to obtain the
ambient gas density at the end of each side of each source and the 
beam power being channeled to each side of each source. The source widths and 
pressures used to derive the ambient gas densities and beam powers 
are presented by \citet{O07}.   
\begin{figure}
\plotone{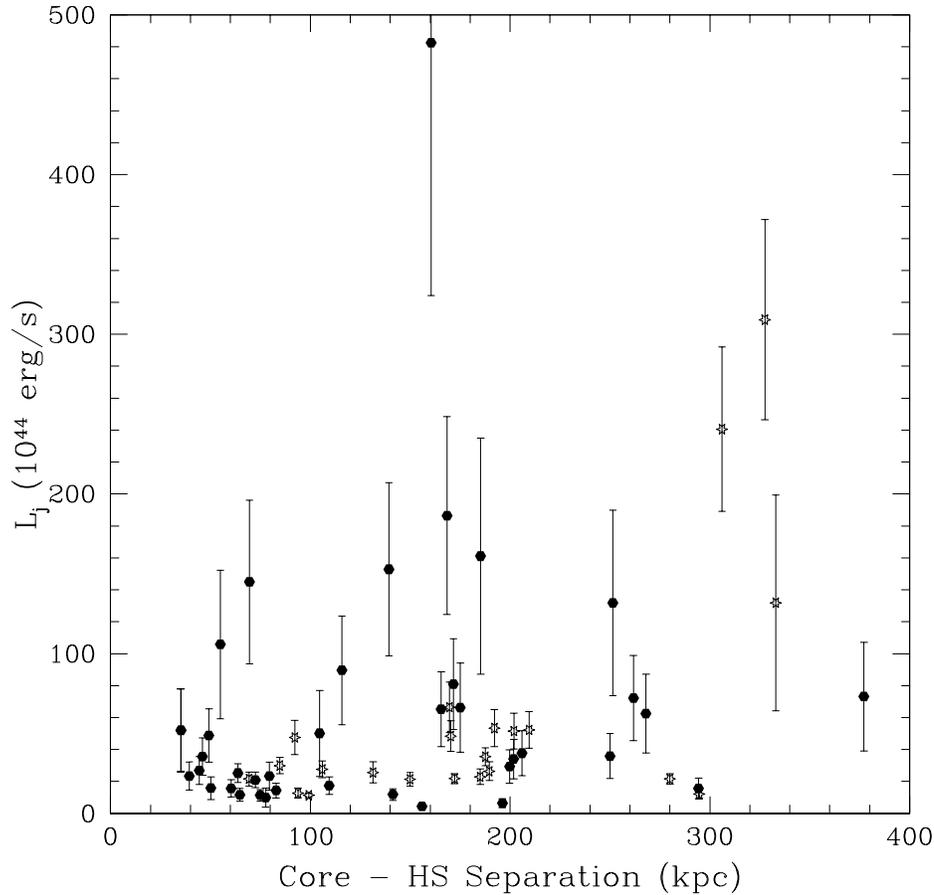}
\caption{As in Figure 1, but for the beam power of each side 
of each source. Note that this beam power is independent 
of whether there are any offsets of the magnetic field strength from 
minimum energy conditions.}
\end{figure}

The composite profile of the 
ambient gas density is shown in Figure 2.
The estimated ambient gas density is quite sensitive to offsets from 
minimum energy conditions, as discussed, for example, by 
\citet{PT91}, \citet{C91}, and \citet{WDW97}.
The values shown here for the velocity, ambient gas density,
beam power, total source lifetime, and total source energy are
obtained assuming the magnetic field strength is one quarter of
the minimum energy value.  This is the value suggested by 
requiring agreement between the ambient gas density indicated
by X-ray measurements and ram pressure confinement of the forward
region of the radio source \citep{PT91,C91,WDW97}.
The ambient gas density
falls with distance from the radio core as expected if the 
sources are close to the center of clusters or protoclusters
of galaxies.
\begin{figure}
\plotone{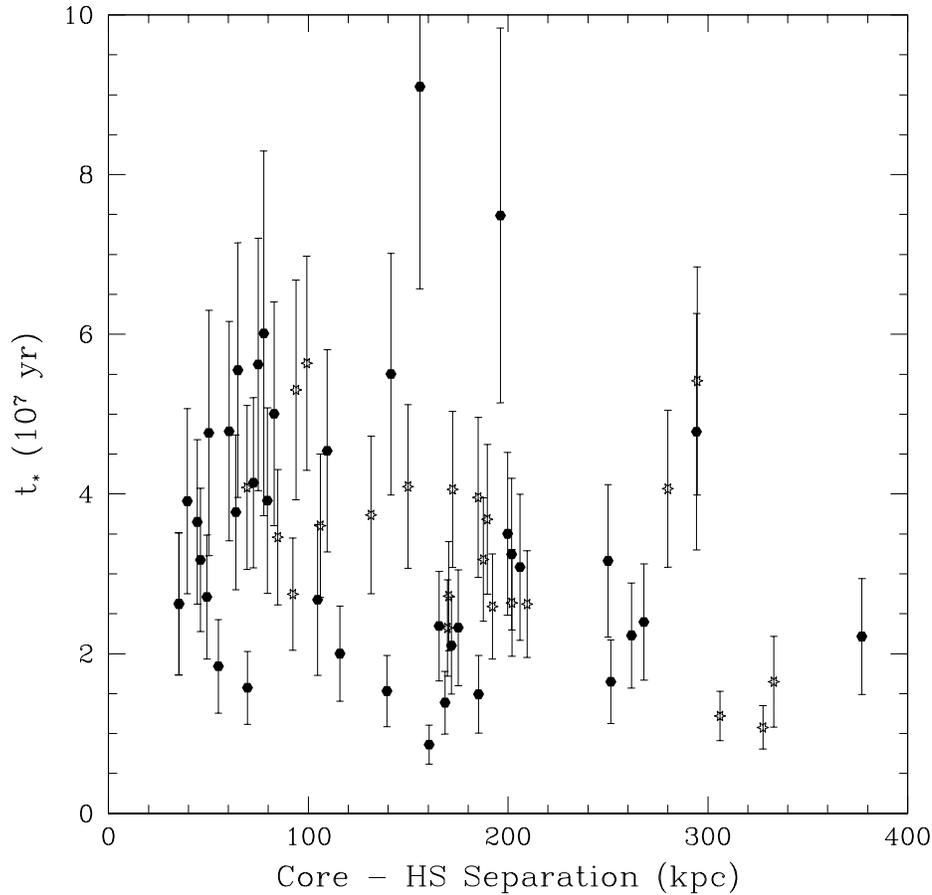}
\caption{As in Figure 1, but for the total lifetime of the 
outflow. }
\end{figure}  

The beam power is plotted as a function of
core-hotspot separation in Figure 3, which
indicates that the beam power of each source is likely
to be roughly constant over the lifetime of the source.
Interestingly, the beam power obtained here 
($L_j \propto vPa^2$) has a very weak dependence on
any offsets of the magnetic field strength from minimum
energy conditions due to a cancellation between the way 
the magnetic field enters $v$ and $P$, 
as demonstrated by \citet{O07}.
Thus, our determination of $L_j$ is essentially independent
of whether the system is close to or quite far from 
minimum energy conditions.

A comparison of the properties of each source with those of
the parent population at similar redshift allow a power-law
relationship to be derived between the beam power of a given
source and the total lifetime of the outflow that powers
the large-scale radio emission, as discussed in detail by \citet{D07} 
for the sample shown here \citep[see also][]{DG02}.  
The total outflow lifetime is shown as
a function of core-hotspot separation in Figure 4. The fact that
the lifetime is independent of core-hotspot separation 
suggests that each source is being randomly sampled during
its lifetime. Typical lifetimes of about $10^7$ years  
are derived. Total source energies are obtained by taking the
product of the beam power and the total source lifetime; typical 
outflow energies of about $10^6 M_{\odot}$ are obtained, as
shown in Figure 5.
\begin{figure}
\plotone{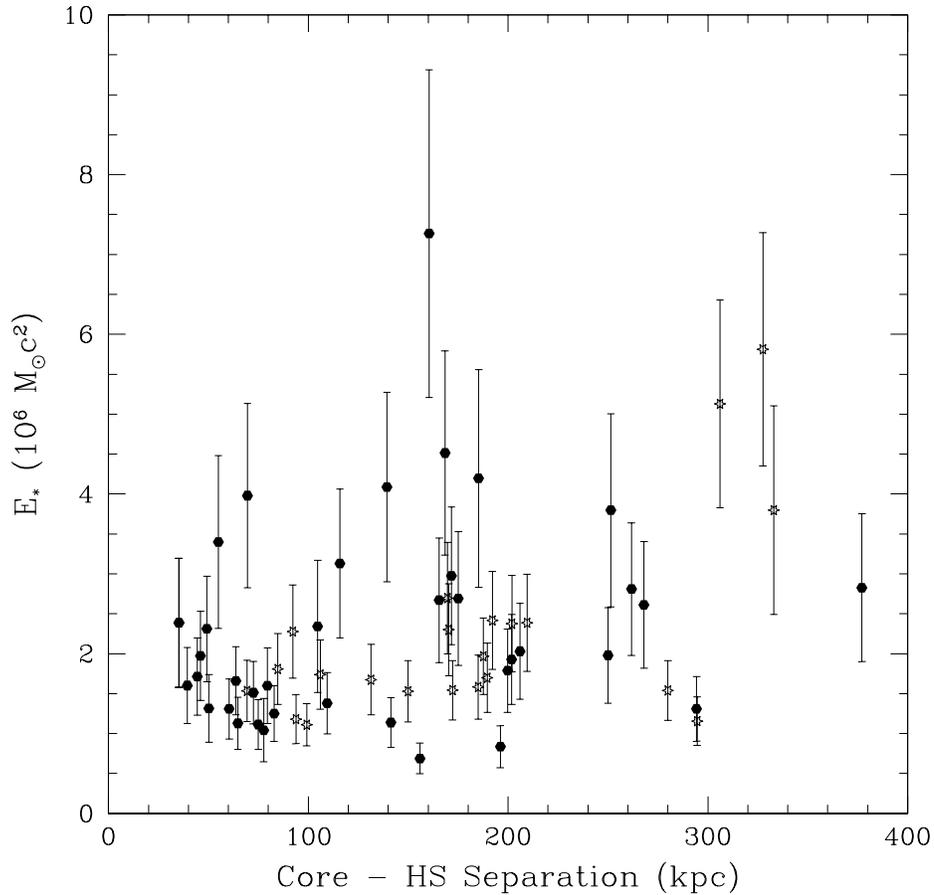}
\caption{As in Figure 1, but for the energy channeled into
the outflow over the total lifetime of the source. }
\end{figure}

\section{Conclusions}

Very powerful classical double radio galaxies provide a
powerful tool to study the properties and environments of
AGN.  These sources have regular radio bridge structures,
indicating that they are propagating at speeds well
into the supersonic regime and that they have negligible
internal backflow speeds (Leahy, Muxlow, and Stephens 1989). This
means that the equations of strong shock physics can be
applied to these systems allowing us to obtain the beam 
power and ambient gas density of each side of each source
from measurements of the source width, pressure, and 
rate of growth of each side of each source. 
In addition, other studies of this type of source indicates a power
law relationship between the beam power and total lifetime of
the source \citep[e.g.][]{DG02}. This allows us to 
obtain an estimate of the total lifetime and outflow
energy of each source.  

\acknowledgements 
We would like to thank the organizers of the conference, 
Travis Rector and Dave De Young, for making this such a successful conference. 
Support for this work was provided in part 
by US National Science Foundation grant number AST-0507465 (R. A. D.). 
This research made use of (1) the NASA/IPAC Extragalactic Database
(NED) which is operated by the Jet Propulsion Laboratory, California
Institute of Technology, under contract with the National Aeronautics and
Space Administration; and (2)  NASA's Astrophysics Data System Abstract
Service. We are grateful to the Penn State University Computer
Center for the use of their IMSL libraries.

\end{document}